\pdfoutput=1
\documentclass{article}
\usepackage{graphicx,amsmath,amssymb}
\usepackage{fullpage}

\newtheorem{theorem}{Theorem}
\newtheorem{lemma}[theorem]{Lemma}
\newtheorem{cor}[theorem]{Corollary}

\newtheorem{openProb}{Open Problem}

\newcommand{\reals}{\ensuremath{\mathbb{R}}}

\newcommand{\freq}{\operatorname{freq}}

\newcommand{\subsubsubsection}[1]{\vspace{10pt} \noindent{\bf{#1}.}}
\newcommand{\subsubsubsectionB}[1]{\noindent{\bf{#1}.}}

\begin{document}

\title{Linear-Space Data Structures for Range Mode Query 
in Arrays\thanks{Work supported in part by 
the Natural Sciences and Engineering Research Council of Canada (NSERC).}}

\author{S.~Durocher\thanks{University of Manitoba, Winnipeg, Canada, 
{\em durocher@cs.umanitoba.ca}}
\and J.~Morrison\thanks{University of Manitoba, Winnipeg, Canada, 
{\em jason\_{}morrison@umanitoba.ca}}}

\date{January 20, 2011}

\maketitle

%%%%%%%%%%%%%%%%%%%%%%%%%%%%%%%%%%%%%%%%%%%%
% ABSTRACT
\begin{abstract}
A mode of a multiset $S$ is an element $a \in S$ of maximum multiplicity;
that is, $a$ occurs at least as frequently as any other element in $S$.
Given a list $A[1:n]$ of $n$ items, we consider the problem of constructing 
a data structure that efficiently answers range mode queries on $A$.
Each query consists of an input pair of indices $(i, j)$ 
for which a mode of $A[i:j]$ must be returned.
We present an $O(n^{2-2\epsilon})$-space static data structure that supports
range mode queries in $O(n^\epsilon)$ time in the worst case,
for any fixed $\epsilon \in [0,1/2]$.
When $\epsilon = 1/2$, this corresponds to
%a linear-space data structure with $O(\sqrt{n})$ worst-case query time.
%This is 
the first linear-space data structure 
to guarantee $O(\sqrt{n})$ query time.
We then describe three additional linear-space data structures
that provide $O(k)$, $O(m)$, and $O(|j-i|)$ query time, respectively,
where $k$ denotes the number of distinct elements in $A$
and $m$ denotes the frequency of the mode of $A$.
%Finally, for any fixed $d$,
%we describe generalizations of our data structures to $d$ dimensions 
%without any increase in asymptotic storage requirements or query time.
%%while maintaining linear space and $O(\sqrt{n})$ query time.
%Finally, we examine generalizing our data structures to $d$ dimensions.
Finally, we examine generalizing our data structures to higher dimensions.
\end{abstract}

%\begin{keyword}
%mode \sep range query \sep data structure \sep linear space \sep array 
%\sep static
%\end{keyword}

%%%%%%%%%%%%%%%%%%%%%%%%%%%%%%%%%%%%%%%%%%%%
% SECTION
\section{Introduction}
\label{sec:intro}

%%%%%%%%%%%%%%%%%%%%%%%%%%%%%%%%%%%%%%%%%%%%
% SUBSECTION
\subsubsubsectionB{Mode and Range Queries}
\label{sec:intro.motivation}
The {\em frequency} of an element $x$ in a multiset $S$, denoted $\freq_S(x)$,
is the number of occurrences (i.e., the multiplicity) of $x$ in $S$.
A {\em mode} of $S$ is an element $a \in S$ such that
%\[ \forall x \in S , \ \freq_S(x) \leq \freq_S(a) . \]
for all $x \in S$, $\freq_S(x) \leq \freq_S(a)$.
A multiset $S$ may have multiple distinct modes;
the frequency of the modes of $S$, denoted by $m$, is unique.

Along with the mean and median of a multiset, the mode is a fundamental 
statistic of data analysis for which efficient computation is necessary.
Given a sequence of $n$ elements ordered in a list $A$, a range query 
seeks to compute the corresponding statistic on the multiset 
determined by a subinterval of the list: $A[i:j]$.
The objective is to preprocess $A$ to construct a data structure
that supports efficient response to one or more subsequent range queries,
where the corresponding input parameters $(i,j)$ are provided at query time.

We assume the RAM model of computation with word size $\Theta(\log u)$,
where elements are drawn from a universe $U = \{ 0 , \ldots , u-1 \}$.
Although the complete set of possible queries can be precomputed and
stored using $\Theta(n^2)$ space, 
practical data structures require less storage 
while still enabling efficient response time.
For all $i$, if $i=j$, then a range query must report $A[i]$.
Consequently, 
% Do we need to say anything about information-theoretic details
% related to compression?
any range query data structure 
for a list of $n$ items requires $\Omega(n)$ storage space in the 
worst case \cite{bose2005}.
This leads to a natural question: how quickly can an $O(n)$-space data
structure answer range queries?
The problem of constructing efficient data structures 
for range median queries has been analyzed extensively
%, including both static and dynamic,
%and both linear- and superlinear-space data structures
\cite{bose2005,brodal2010,brodal2009,chan2010,gagie2009,gfeller2009,%
har-peled2008,jorgensen2010,jorgensen2011,krizanc2005,petersen2008,%
petersen2009}.
%In the discrete setting,
A range mean query is equivalent to a normalized range sum query 
(partial sum query), 
for which a precomputed prefix-sum array provides a linear-space
static data structure with constant query time
\cite{krizanc2005}.
%In this paper we examine
%linear-space static data structures for the range mode query problem.
As expressed recently by Brodal et al.\ regarding the current status of
the range mode query problem:
``The problem of finding the most frequent element within a given array
range is still rather open.'' \cite[page 2]{brodal2010}.
See Section~\ref{sec:relatedWork.rangeMode} for an overview of
the current state of the range mode query problem.

%%%%%%%%%%%%%%%%%%%%%%%%%%%%%%%%%%%%%%%%%%%%
% SUBSECTION
\subsubsubsection{Our Results}
\label{sec:intro.results}
%We begin with a brief discussion of related work in 
%Section~\ref{sec:relatedWork}.
%Our main contribution is the definition and analysis of new data structures
%and their corresponding range mode query algorithms
%in Section~\ref{sec:dataStructures}.
Given an array $A[1:n]$ of $n$ items, 
we present an $O(n^{2-2\epsilon})$-space static data structure that supports
range mode queries in $O(n^\epsilon)$ time in the worst case,
for any fixed $\epsilon \in [0,1/2]$.
When $\epsilon = 1/2$, this corresponds to
the first linear-space data structure to guarantee $O(\sqrt{n})$ query time.
Prior to our work, the previous fastest linear-space data structure
by Krizanc et al.\ \cite{krizanc2005}
supported range mode queries in $O(\sqrt{n}\log\log n)$ time;
our data structure borrows ideas developed by Krizanc et al.\
and augments their data structure to eliminate dependence on predecessor
queries (see Proposition~\ref{lem:minFreqQuery}).
We describe three additional $O(n)$-space data structures that 
provide $O(k)$, $O(m)$, and $O(|j-i|)$ query time, respectively,
where $k$ denotes the number of distinct elements in $A$.
%In Section~\ref{sec:higherDim}
Finally we discuss generalizations of our data structures 
to $d$ dimensions for any fixed $d$.
%%while maintaining linear space and $O(\sqrt{n})$ query time.
To the authors' knowledge, this is the first examination of 
multidimensional range mode query.
%Section~\ref{sec:discussion}
%concludes with a discussion of directions for future research.

%%%%%%%%%%%%%%%%%%%%%%%%%%%%%%%%%%%%%%%%%%%%
% SECTION
\section{Related Work}
\label{sec:relatedWork}

%%%%%%%%%%%%%%%%%%%%%%%%%%%%%%%%%%%%%%%%%%%%
% SUBSECTION
\subsubsubsectionB{Computing a Mode}
\label{sec:relatedWork.mode}
The mode of a multiset $S$ of $n$ items can be found in $O(n \log n)$ time
by sorting $S$ and scanning the sorted list to identify the longest sequence
of identical items.
Due to the corresponding lower bound on the worst-case time
for solving the element uniqueness problem,
finding a mode requires $\Omega(n \log n)$ time in the worst case;
that is, the decision problem of determining whether $m > 1$ requires 
$\Omega(n \log n)$ time in the worst case \cite{skiena2008}.
Better bounds on the worst-case time
are obtained by parameterizing in terms of $m$ or $k$.
A worst-case time of $O(n \log k)$ is easily achieved by 
inserting the $n$ elements into a balanced search tree 
in which each node stores a key and its frequency.
%Each time a frequency is incremented, its value is compared against
%the multiplicity of the current mode. 
Munro and Spira \cite{munro1976} 
describe an $O(n \log (n/m))$-time algorithm 
for finding a mode and a corresponding lower bound of
$\Omega(n \log (n/m))$ on the worst-case time.

If distinct elements in $S$ can be mapped efficiently (i.e., in constant time)
to distinct integers in
the range $\{1, \ldots, k'\}$, for some $k'$, then a mode of $S$ 
can be found in $O(n+k')$ time using $O(n+k')$ space.
This is achieved by identifying a maximum element in a frequency table for $S$
of size $k'$.
This method is analogous to counting sort.
%, with the exception that only 
%the largest value (most frequent element) is returned.
A similar algorithm for computing a mode can be implemented using hash tables.

We include the following lemma to which we refer in 
Section~\ref{sec:dataStructure1}:

\begin{lemma}[Krizanc et al.\ \cite{krizanc2005}]
\label{lem:krizanc}
Let $A$ and $B$ be any multisets.
If $c$ is a mode of $A \cup B$ and $c \not\in A$, then $c$ is a mode of $B$.
\end{lemma}

%%%%%%%%%%%%%%%%%%%%%%%%%%%%%%%%%%%%%%%%%%%%
% SUBSECTION
\subsubsubsectionB{Range Mode Query}
\label{sec:relatedWork.rangeMode}
%The one-sided range mode query problem is straightforward:
%a precomputed array $B$ of size $n$ can store a mode of $A[1:i]$ in $B[i]$
%to provide constant-time queries.
Naturally, a mode of the query interval $A[i:j]$
can be computed directly without preprocessing 
using any of the methods described in Section~\ref{sec:relatedWork.mode}.
% $O((j-i) \log [(j-i)/m'])$ time using the algorithm 
%of Munro and Spira \cite{munro1976},
%where $m'$ denotes the frequency of the modes of $A[i:j]$ 
%(which is at most $m$).
Krizanc et al.\ \cite{krizanc2005} describe data structures that
provide constant-time queries using $O(n^2 \log\log n/ \log n)$ space
and $O(n^\epsilon \log n)$-time queries using $O(n^{2-2\epsilon})$ space,
for any fixed $\epsilon \in (0, 1/2]$.
Petersen and Grabowski \cite{petersen2009} 
improve the first bound to constant time and $O(n^2 \log\log n/ \log^2 n)$ space
and Petersen \cite{petersen2008} improves the second bound to
$O(n^\epsilon)$-time queries using $O(n^{2-2\epsilon})$ space,
for any fixed $\epsilon \in [0, 1/2)$.
When $\epsilon = 1/2$,
the data structure of Krizanc et al.\ \cite{krizanc2005}
requires only linear space and provides $O(\sqrt{n}\log\log n)$ query time.
Although its space requirement is almost linear in $n$ as $\epsilon$
approaches $1/2$, 
the data structure of Petersen \cite{petersen2008} requires $\omega(n)$ space.
Furthermore, the construction becomes impractical 
as $\epsilon$ approaches $1/2$ 
(the number of levels in a hierarchical set of tables and hash functions 
approaches $\infty$ as $\epsilon \to 1/2$)
and no obvious modification reduces its space requirement to $O(n)$.
Greve et al.\ \cite{greve2010}
prove a lower bound of $\Omega(\log n / \log(s \cdot w/n))$ 
query time for any data structure that uses $s$ memory cells of $w$ bits.

Bose et al.\ \cite{bose2005} consider approximate range mode queries,
in which the objective is to return an element whose frequency is
at least $\alpha \cdot m$.
They give a data structure that requires $O(n/(1-\alpha))$ space
and answers approximate range mode queries in $O(\log\log_{1/\alpha}n)$ time
for any fixed $\alpha \in (0,1)$,
as well as data structures that provide constant-time queries for 
$\alpha \in \{1/2, 1/3, 1/4\}$, using space $O(n\log n)$,
$O(n\log\log n)$, and $O(n)$, respectively.
Greve et al.\ \cite{greve2010} give a linear-space data structure
that supports approximate range mode queries in constant time 
for $\alpha = 1/3$,
and an $O(n \cdot \alpha / (1-\alpha))$-space data structure that supports 
approximate range mode queries in $O(\log(\alpha/(1-\alpha)))$ time
for any fixed $\alpha \in [1/2,1)$.
%and an $O(n / \epsilon)$-space data structure that supports 
%approximate range mode queries in $O(\log(1/\epsilon))$ time
%for any fixed $\epsilon \in (0,1]$, where $\alpha = 1/(1+\epsilon)$.

%%%%%%%%%%%%%%%%%%%%%%%%%%%%%%%%%%%%%%%%%%%%
% SUBSECTION
\subsubsubsection{Continuous Space versus Array Input}
\label{sec:relatedWork.continuous}
A vast literature studies the problems of %multidimensional 
geometric range searching in continuous Euclidean space;
that is, data points are positioned arbitrarily in $\reals^d$.
See the survey by Agarwal \cite{agarwal2004} for an overview of results.
The range query problems considered in this paper, however, 
restrict attention to array input.
Although a range query on an array can be viewed as a restricted case
of a more general range searching problem 
(e.g., a point set with regular spacing),
the algorithmic techniques differ greatly between the two settings
when $d \geq 2$.
%solution techniques tend to differ extensively for
%range searching problems set in continuous Euclidean space
%versus those restricted to array input.
%In particular, 
%data points within the query range are immediately accessible in an array.
When $d=1$, however, a geometric range mode query problem reduces 
to array range mode query.
In particular, 
the rank of each data point in Euclidean space corresponds to its array index.
It suffices to compute the ranks of the 
respective successor and predecessor
of the endpoints of the query interval to identify the indices $i$ and $j$,
and to return the corresponding array range mode query on $A[i:j]$.

In addition to results on the median, mode, and sum range query problems 
discussed in Sections~\ref{sec:intro.motivation}
and~\ref{sec:relatedWork.rangeMode},
other range query problems examined on arrays include
semigroups \cite{alon1987,yao1982,yao1985}, 
extrema (e.g., range minimum or maximum) 
\cite{bender2000,berkman1989,demaine2009,fischer2006,fischer2007,fischer2010,%
fischer2010b,gabow1984},
selection or quantiles (for which the median is a special case) 
\cite{gagie2009,gfeller2009,jorgensen2010,jorgensen2011},
dominance or rank (counting the number of elements in the query range 
that exceed a given input threshold) \cite{jaja2004,jorgensen2010},
coloured range (counting/enumerating the distinct elements in
the query range) \cite{gagie2009},
and $k$-frequency (determining whether any element has frequency $k$)
\cite{greve2010}.
Recently, range query problems have been examined on multidimensional
arrays, including partial sums \cite{chazelle1989},
range minimum \cite{amir2007,brodal2010b,demaine2009,poon2003,yuan2010},
median \cite{gfeller2009}, and selection \cite{gagie2009}.
%%To the authors' knowledge, this paper contains the first examination of 
%%multidimensional range mode query.

%%%%%%%%%%%%%%%%%%%%%%%%%%%%%%%%%%%%%%%%%%%%
% SECTION
\section{Sparse Mode Table Method: 
$O(n^\epsilon)$ Query Time and $O(n^{2-2\epsilon})$ Space}
\label{sec:dataStructure1}

%Given an array $A[1:n]$ of $n$ items, 
%we describe three $O(n)$-space static data structures that answer
%range mode queries in $O(\sqrt{n})$, $O(k)$, and $O(|j-i|)$ time 
%in the worst case, respectively,
%where $k$ denotes the number of distinct elements in $A$.
%These can be combined into a single linear-space data structure that 
%provides $O(\min\{ \sqrt{n}, k, |j-i| \})$ worst-case query time
%by simply accessing the data structure whose query time corresponds 
%to the minimum of $\{ \sqrt{n}, k, |j-i| \}$.

\noindent
In the worst case, for every range mode query processed, 
the data structure of Krizanc et al.\ \cite{krizanc2005}
makes a sequence of $\Theta(n^\epsilon)$ predecessor queries,
each requiring $\Theta(\log\log n)$ time, for a total query time of 
$\Theta(n^\epsilon\log\log n)$.
%As noted by Krizanc et al.,
%this time could be reduced to $O(\sqrt{n} \log\log n)$ by using a
%van Emde Boas tree \cite{vanEmdeBoas1975,vanEmdeBoas1976,vanEmdeBoas1977}
%or a y-fast trie \cite{willard1983}
%for predecessor search.
We build on the data structure of Krizanc et al.\
and introduce a different technique that avoids predecessor search entirely.
Section~\ref{sec:dataStructure1} establishes the following theorem
and the corresponding corollary that follows when $\epsilon = 1/2$:

\begin{theorem}
\label{thm:mainResult}
Given an array $A[1:n]$ of $n$ items, for any $\epsilon \in [0, 1/2]$
there exists a data structure requiring 
$O(n^{2-2\epsilon})$ storage space that supports range mode queries on $A$ in 
$O(n^\epsilon)$ time in the worst case. 
\end{theorem}

\begin{cor}
Given an array $A[1:n]$ of $n$ items, there exists a data structure requiring 
$O(n)$ storage space that supports range mode queries on $A$ in 
$O(\sqrt{n})$ time in the worst case. 
\end{cor}

%%%%%%%%%%%%%%%%%%%%%%%%%%%%%%%%%%%%%%%%%%%%
% SUBSECTION
\subsubsubsectionB{Data Structure Precomputation}
\label{sec:dataStructure1.precomp}
Suppose the elements of $A[1:n]$ are drawn from an ordered bounded universe $U$.
Let $D = \{a_1, \ldots , a_k\} \subseteq U$ denote the set of distinct elements
stored in $A$.
Construct an array $B[1:n]$ 
such that for each $i$, $B[i]$ stores the rank of $A[i]$ in $D$.
Therefore, $B[i] \in \{ 1, \ldots, k \}$.
For any $a$, $i$, and $j$,
$B[a]$ is a mode of $B[i:j]$ if and only if $A[a]$ is a mode of $A[i:j]$.
Performing computation on array $B$ instead of array $A$
allows direct array referencing using the values stored in $B$ as indices.
%The data structures described in 
%Sections~\ref{sec:dataStructures.rootN}
%through~\ref{sec:dataStructures.j-i}
%assume precomputation of array $B$ and the value $k = |D|$.
For simplicity, we describe our data structures in terms of array $B$;
a table look-up provides a direct bijective mapping 
from $\{1, \ldots, k\}$ to $D$.
Set $D$, array $B$, and the value $k$
are independent of any query range and
can be computed in $O(n \log k)$ time during preprocessing.
% by sorting $A$ and scanning the sorted array.
%The value $k$, a mode $m$, and its frequency are straightforward
%to compute by scanning the sorted array.

%Our data structures contain frequency tables.
Given fixed $a$ and $b$, 
array $C[1:k]$ is a {\em frequency table} for $B[a:b]$ if, for each $i$, 
$C[i]$ stores the number of occurrences of element $i$ in $B[a:b]$.
%Thus, $C[B[i]]$ corresponds to the frequency of $A[i]$ in $A[a:b]$.
For any $j > i$, if $C_i[1:k]$ is a frequency table for $B[1:i]$ and
$C_j[1:k]$ is a frequency table for $B[1:j]$, 
then for each $x$, $C_j[x]-C_i[x]$ is the frequency of $B[x]$ in $B[i+1:j]$.

For each $a \in \{1, \ldots , k\}$, let $Q_a = \{ b \mid B[b] = a\}$.
That is, $Q_a$ is the set of indices $b$ such that $B[b] = a$.
For any $a$,
a range counting query for element $a$ in $B[i:j]$ can be answered by 
searching for the predecessors of $i$ and $j$, respectively, in the set $Q_a$;
the difference of the indices of the two predecessors is the frequency of $a$
in $B[i:j]$ \cite{krizanc2005}.
As noted above, implementing such a range counting query 
using an efficient predecessor data structure 
requires $\Theta(\log\log n)$ time in the worst case.

The following related decision problem, however, 
can be answered in constant time by a linear-space data structure:
does $B[i:j]$ contain at least $q$ instances of element $B[i]$?
This question can be answered by a select query that returns the index of the
$q$th instance of $B[i]$ in $B[i:n]$.
For each $a \in \{1, \ldots , k\}$, 
store the set $Q_a$ as an ordered array (also denoted $Q_a$ for simplicity).
Define a rank array $B'[1:n]$ such that for all $b$, 
$B'[b]$ denotes the rank of $B[b]$ in $B[1:n]$
(i.e., the index of $b$ in $Q_{B[b]}$). 
%Define an array $B'[1:n]$ such that for all $a$ and $b$, 
%if $B[b] = a$, then $B'[b]$ denotes the index (i.e., the rank) of $b$ in $Q_a$.
%Array $B'$ serves as an inverse of $Q_a$; 
%thus, if $B[b] = a$, then $Q_a[B'[b]] = b$.
%Array $B'$ serves as an inverse of $Q_{B[b]}$; 
%thus, $Q_{B[b]}[B'[b]] = b$.
Given any $q$, $i$, and $j$,
to determine whether $B[i:j]$ contains at least $q$ instances of $B[i]$
it suffices to check whether $Q_{B[i]}[B'[i]+q-1] \leq j$.
Since array $Q_{B[i]}$ stores the sequence of indices of instances of element
$B[i]$ in $B$, 
looking ahead $q-1$ positions in $Q_{B[i]}$ returns
the index of the $q$th occurrence of element $B[i]$ in $B[i:n]$;
if this index is at most $j$, then the frequency of $B[i]$ in $B[i:j]$ is at
least $q$.
If the index $B'[i]+q-1$ exceeds the size of the array $Q_{B[i]}$, 
then the query returns a negative answer.
This gives the following lemma:

\begin{lemma}
\label{lem:minFreqQuery}
Given an array $A[1:n]$ of $n$ items, 
there exists a data structure requiring $O(n)$ storage space 
that can determine in constant time 
for any $\{i, j\} \subseteq \{1 , \ldots, n\}$ and any $q$
%for any $0 \leq i \leq j \leq n-1$ and any $q$
whether $A[i:j]$ contains at least $q$ instances of element $A[i]$.
\end{lemma}

Following Krizanc et al.\ \cite{krizanc2005}, 
given any $\epsilon \in [0,1/2]$
we partition array $B$ into $t$ blocks of size 
$s = \lceil n^\epsilon \rceil$, 
where $t = \lceil n/s \rceil \leq \lceil n^{1-\epsilon} \rceil$.
That is, for each $i \in \{ 0, \ldots, t - 2 \}$,
the $i$th block spans $B[i\cdot s + 1:(i+1)s]$
and the last block spans $B[(t-1)\cdot s + 1 : n]$.
We precompute tables $S[0:t-1,0:t-1]$ and $S'[0:t-1,0:t-1]$,
each of size $\Theta(t^2)$,
such that for any $\{b_i, b_j\} \subseteq \{0, \ldots, t-1\}$,
$S[b_i,b_j]$ stores a mode of $B[b_i\cdot s+1 : (b_j+1)s]$
and $S'[b_i,b_j]$ stores the corresponding frequency.

Finally, we need a frequency table $C[1:k]$ of size $k$, initialized to zero.
The arrays $Q_1, \ldots, Q_k$ can be constructed in $O(n)$ total time
in a single scan of array $B$.
The arrays $S$ and $S'$ can be constructed in $O(n^{2-\epsilon})$ time by 
scanning array $B$ $t$ times, computing one row of each array 
$S$ and $S'$ per scan.
Thus, the total precomputation time required to initialize 
the data structure is $O(n^{2-\epsilon})$.

\begin{figure}
\centering
\includegraphics[width=0.7\linewidth]{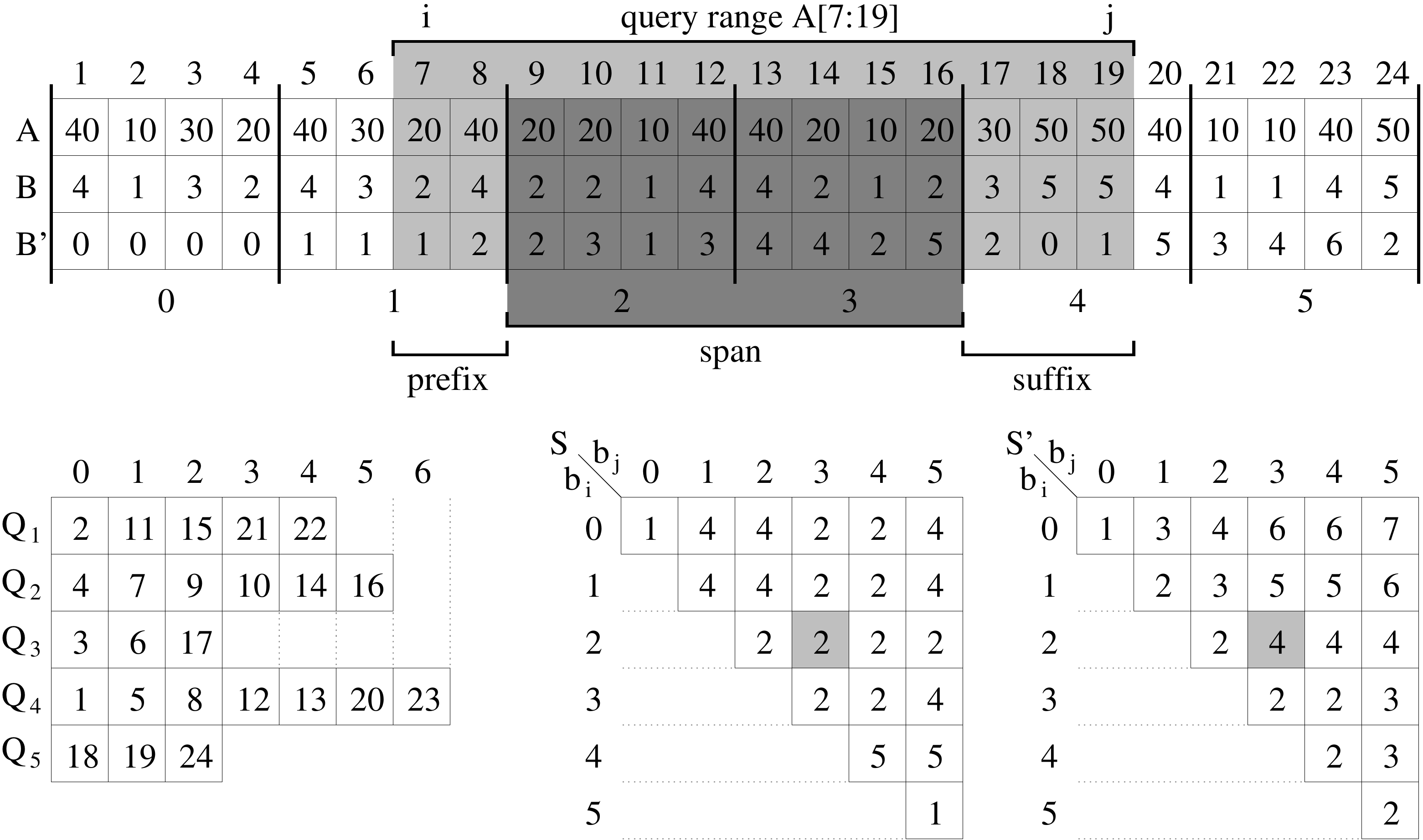}
\caption{{\bf Example of the sparse mode table method data structure.}
The number of list items is $n=24$, of which $k=5$ are distinct.
If $\epsilon = 3/8$, 
the array is partitioned into $t= \lceil n / s \rceil = 6$ 
blocks of size $s= \lceil n^\epsilon \rceil = 4$. 
The query range is
$A[i:j]=A[7:19]$, for which the unique mode is 20, occurring with frequency 5. 
The corresponding mode of $B[i:j]$ is 2. 
The query range $B[7:19]$ is partitioned into 
the prefix $B[7:8]$, the span $B[9:16]$, and the suffix $B[17:19]$.
The span covers blocks $b_i=2$ to $b_j=3$,
for which the corresponding mode is $S[2,3] = 2$, 
occurring with frequency $S'[2,3]= 4$.}
\label{fig:dataStructure1}
\end{figure}

%%%%%%%%%%%%%%%%%%%%%%%%%%%%%%%%%%%%%%%%%%%%
% SUBSECTION
\subsubsubsection{Range Mode Query Algorithm}
Given a query range $B[i:j]$,
let $b_i = \lceil (i-1) / s \rceil$ and $b_j = \lfloor j / s \rfloor - 1$
denote the respective indices of the first and last blocks  
completely contained within $B[i:j]$.
We refer to $B[b_i\cdot s+1:(b_j+1)s]$ as the {\em span} of the query range,
to $B[i:\min\{b_i\cdot s, j\}]$ as its {\em prefix},
and to $B[\max\{(b_j+1)s+1, i\}:j]$ as its {\em suffix}.
One or more of the prefix, span, and suffix may be empty;
in particular, if $b_i > b_j$, then the span is empty.
See the example in Figure~\ref{fig:dataStructure1}.

The value $c = S[b_i,b_j]$ is a mode of the span
with corresponding frequency $f_c = S'[b_i,b_j]$. 
If the span is empty, then let $f_c = 0$.
By Lemma~\ref{lem:krizanc}, either $c$ is a mode of $B[i:j]$ 
or some element of the prefix or suffix is a mode of $B[i:j]$.
Thus, to identify a mode of $B[i:j]$,
we verify for every element in the prefix and suffix
whether its frequency in $B[i:j]$ exceeds $f_c$ and,
if so, we identify this element as a {\em candidate} mode and
count its additional occurrences in $B[i:j]$.
We present the details of this procedure for the prefix;
an analogous procedure is applied to the suffix.

We now describe how 
to compute the frequency of all candidate elements in the prefix over 
the range $B[i:j]$, storing these values in the frequency table $C$.
Sequentially scan the items in the prefix starting at the leftmost index, $i$,
and let $x$ denote the index of current item.
If $C[B[x]] > 0$, then an instance of element $B[x]$ appears in $B[i:x-1]$, 
and its frequency has been counted already; 
in this case, simply skip $B[x]$ and increment $x$.
If $C[B[x]] = 0$, check whether $Q_{B[x]}[B'[x]+f_c-1] \leq j$
(i.e., verify whether $B[x]$ is a candidate).
If so, then the frequency of $B[x]$ in $B[i:j]$ is at least $f_c$.
The exact frequency of $B[x]$ in $B[i:j]$ can be counted by
a linear scan of $Q_{B[x]}$, starting at index $B'[x]+f_c-1$
and terminating upon reaching either an index $y$ such that $Q_{B[x]}[y] > j$ 
or the end of array $Q_{B[x]}$ (i.e., $y = |Q_{B[x]}| + 1$).
That is, $Q_{B[x]}[y]$ denotes the index of the first instance of element
$B[x]$ that lies beyond the query range $B[i:j]$ (or no such element exists).
Consequently, the frequency of $B[x]$ in $B[i:j]$ is $y - B'[x]$.
Store this value in $C[B[x]]$.

%\begin{prop}
%\label{prop:prefixTime}
%Given an array $A[1:n]$ of $n$ items, 
%there exists a data structure requiring $O(t^2 + n)$ storage space that
%can determine in $O(s)$ time for any $\{i,j\} \subseteq \{0, \ldots, n-1\}$
%whether $S[i,j]$ i
%The total time required to 
%\end{prop}

An analogous procedure is repeated for the suffix.
Upon completing the scans of the prefix and suffix, 
we identify a maximum value in array $C$;
its index corresponds to a mode of $B[i:j]$.
Only non-zero entries in $C$ need be examined (and subsequently reset to zero);
this is achieved by making a second scan of the prefix and suffix
and examining the corresponding elements in array $C$.

%%%%%%%%%%%%%%%%%%%%%%%%%%%%%%%%%%%%%%%%%%%%
% SUBSECTION
\subsubsubsection{Storage Space and Query Time}
If the prefix and suffix are empty, then $S[b_i,b_j]$ is a mode of $B[i:j]$,
and this value is returned in constant time.
Without loss of generality, suppose the prefix contains at least one item.
Consider an arbitrary index $x \in \{i, \ldots, b_i\cdot s-1\}$ 
during the scan of the prefix.
If $C[B[x]] > 0$, then $B[x]$ is processed in constant time.
Therefore, suppose $C[B[x]] = 0$.
That is, $x$ corresponds to the index of the first instance of $B[x]$ in
the prefix.
Consequently, the frequency of $B[x]$ in $B[i:j]$, denoted $f_x$,
is equal to its frequency in $B[x:j]$.
By Lemma~\ref{lem:minFreqQuery},
determining whether $f_x \geq f_c$ requires only constant time.
Any item $B[x]$ that is not a candidate is processed in constant time. 
Therefore, suppose $B[x]$ is a candidate.
Since the prefix and suffix each have size at most $s-1$,
$f_c \leq f_x \leq 2(s-1)$.

Item $B[x]$ incurs a cost of $O(f_x-f_c)$ time for its first occurrence, 
and $O(1)$ time for subsequent occurrences. 
Since $f_c$ is the frequency of the mode of the span,
at least $f_x - f_c$ instances of $B[x]$ must occur in the prefix or suffix.
In other words, instances of element $B[x]$ incur a total cost of $O(c_x)$ time,
where $c_x$ denotes the frequency of $B[x]$ in the prefix and suffix.
Since the number of items in the prefix and suffix is at most $2(s-1)$, 
the total cost for processing the prefix is $O(s)$.
By an analogous argument, the total cost for processing the suffix 
is also $O(s)$.
Identifying the maximum element in array $C$ and re-initializing $C$ to zero
requires $O(s)$ time.
Therefore, a range mode query requires $O(s) = O(n^\epsilon)$ 
time in the worst case.
The data structure requires 
$O(n)$ space to store the arrays $A$, $B$, and $B'$,
$O(n)$ total space to store the arrays $Q_1, \ldots, Q_k$,
and $O(t^2) = O(n^{2-2\epsilon})$ space to store the tables $S$ and $S'$.
This gives $O(n^{2-2\epsilon})$ total space for $O(n^\epsilon)$ worst-case
query time for any $\epsilon \in [0,1/2]$, proving Theorem~\ref{thm:mainResult}.
As mentioned earlier, $\Omega(n)$ space is required.
Therefore, increasing $\epsilon$ beyond $1/2$
increases query time without decreasing space.

%%%%%%%%%%%%%%%%%%%%%%%%%%%%%%%%%%%%%%%%%%%%
% SECTION
\section{Additional Linear-Space Range Mode Query Data Structures}
\label{sec:dataStructures2}

\noindent
%We discuss three three additional $O(n)$-space data structures that 
%provide $O(k)$, $O(m)$, and $O(|j-i|)$ query time, respectively,
%where $k$ denotes the number of distinct elements in the array
%and $m$ denotes the frequency of its mode.
We apply results from Section~\ref{sec:dataStructure1} to obtain 
three additional $O(n)$-space data structures, giving the following 
theorem:

\begin{theorem}
\label{thm:otherResults}
Given an array $A[1:n]$ of $n$ items, 
there exists a data structure requiring 
$O(n)$ storage space that supports range mode queries on any $A[i:j]$ in 
$O(\min\{ \sqrt{n}, k, |j-i|, m + \log\log n \})$ time in the worst case,
where $k$ denotes the number of distinct elements in $A$
and $m$ denotes the frequency of the mode of $A$.
\end{theorem}

%%%%%%%%%%%%%%%%%%%%%%%%%%%%%%%%%%%%%%%%%%%%
% SUBSECTION
\subsection{Sparse Frequency Table Method: $O(k)$ Query Time 
and $O(n)$ Space}
\label{sec:dataStructures.k}
\noindent
We now describe an $O(k+s)$ query time and $O(n + n \cdot k/s)$-space
data structure for any fixed $s \in [1,n]$.
When $s \in\Theta(k)$, our data structure requires $O(n)$ space 
and supports range mode queries in $O(k)$ time.
A value of $s \in o(k)$ (respectively, $s \in \omega(k)$) 
results in $\omega(n)$ space ($\omega(k)$ time) without any reduction in
query time (space).

\begin{figure}
\centering
\includegraphics[width=0.6\linewidth]{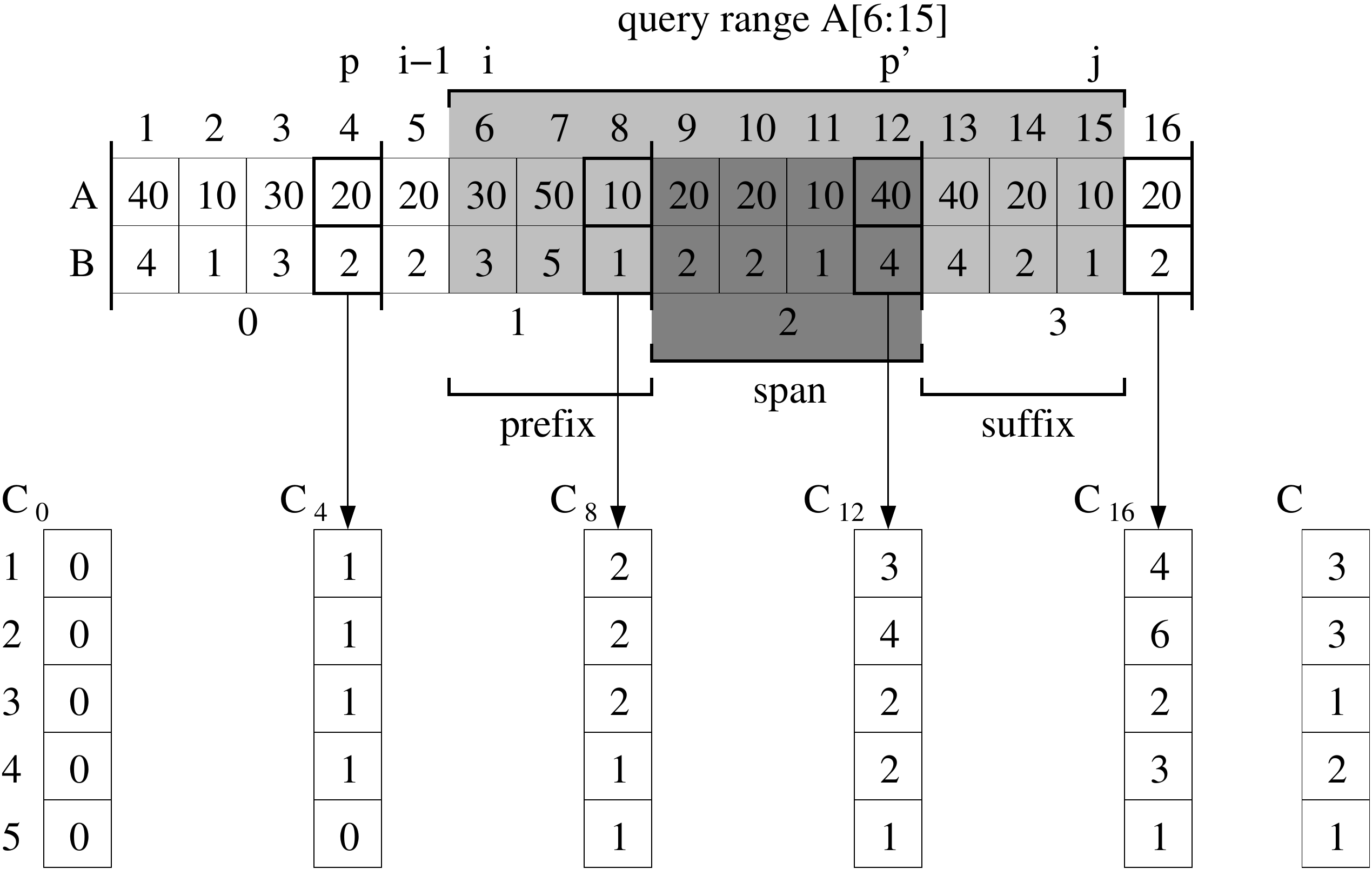}
\caption{{\bf Example of the sparse frequency table method data structure.}
The number of list items is $n=16$, of which $k=5$ are distinct. 
The array is partitioned into four blocks of size $s=4$.
The query range is $A[i:j]=A[6:15]$, for which elements 10 and 20 are 
modes, each occurring with frequency 3. 
The corresponding modes of $B[i:j]$ are 1 and 2.
Thus, $C[1] = C[2] = 3$ is the maximum value in the frequency array $C$.}
\label{fig:linSpaceDataStructure}
\end{figure}

%%%%%%%%%%%%%%%%%%%%%%%%%%%%%%%%%%%%%%%%%%%%
\subsubsubsection{Data Structure Precomputation}
For each $p \in \{1, \ldots, n\}$ such that $p \bmod s = 0$, 
construct a frequency table $C_p[1:k]$ for the range $B[1:p]$.
Create one additional array $C_0[1:k]$, initialized to zero.
There are $\lceil n/s \rceil + 1$ such arrays $C_i$.
See Figure~\ref{fig:linSpaceDataStructure}.
The preprocessing time required is $O(n + n \cdot k/s)$ 
(or $O(n\log k + n \cdot k/s)$ time if $k$ or $B$ must be computed).

%%%%%%%%%%%%%%%%%%%%%%%%%%%%%%%%%%%%%%%%%%%%
\subsubsubsection{Range Mode Query Algorithm}
% possibly reconcile variable names: 
% b_i and b_j -> p and q
Array $B$ is partitioned into blocks of size $s$ 
as in Section~\ref{sec:dataStructure1}.
Given a query range $B[i:j]$,
we refer to the sequence of blocks completely covered by $B[i:j]$ as
the span, and to the remaining subarrays as the prefix and suffix, respectively.
A query on $B[i:j]$ is performed as follows:
\begin{enumerate}
\item 
%Let $p = (i-1) - [(i-1) \bmod s ]$ and 
Let $p = s \lfloor (i-1) / s \rfloor$ and 
%let $p' = j - [j \bmod s ]$.
let $p' = s \lfloor j/s \rfloor$.
That is, $p$ is the largest $p \leq i-1$ such that array $C_p$ is defined.
Similarly, $p'$ is the largest $p' \leq j$ such that array $C_{p'}$ is defined.
\item 
Create an array $C[1:k]$ such that
for each $x$, $C[x] \leftarrow C_{p'}[x] - C_p[x]$.
Upon completing this step,
$C$ is a frequency table for the span $B[p+1:p']$.
\item 
For each $x \in \{p + 1, \ldots, i-1\}$,
set $C[B[x]] \leftarrow C[B[x]] - 1$.
For each $x \in \{p' + 1, \ldots, j\}$,
set $C[B[x]] \leftarrow C[B[x]] + 1$.
Upon completing this step
$C$ is a frequency table for the entire query range $B[i:j]$.
\item 
Find a maximum value in $C$.
If $x'$ is an index that maximizes $C[x']$, then $B[x']$ is a mode of $B[i:j]$.
\end{enumerate}

%%%%%%%%%%%%%%%%%%%%%%%%%%%%%%%%%%%%%%%%%%%%
\subsubsubsectionB{Storage Space and Query Time}
The data structure consists of arrays $A$ and $B$, requiring $O(n)$ space,
and $O\lceil n/s \rceil + 1$ frequency tables of size $k$.
Thus, the total space required by the data structure is $O(n + n \cdot k/s)$.
Steps 1 through 4 of the algorithm require $O(1)$, $O(k)$, $O(s)$, and $O(k)$ 
time, respectively.
This gives $O(n + n \cdot k/s)$ total space for $O(k+s)$ query time.
%Therefore, when $s \in \Theta(k)$, this corresponds to 
%$O(n)$ space and $O(k)$ query time.

%%%%%%%%%%%%%%%%%%%%%%%%%%%%%%%%%%%%%%%%%%%%
% SUBSECTION
\subsection{Low Frequency Mode Method: 
$O(m + \log\log n)$ Query Time and $O(n)$ Space}
\label{sec:dataStructures.lowM}
\noindent
Using a combination of ideas 
from Section~\ref{sec:dataStructure1}
and from an approximate range mode query data structure of 
Greve et al.\ \cite{greve2010},
we briefly describe a range mode data structure parameterized in terms of 
the frequency of the mode, $m$, with good bounds on space and query time
when $m$ is small (e.g., $m \in O(\sqrt{n})$).

As in Section~\ref{sec:dataStructure1},
the rank array $B'$ and the arrays $Q_1,\ldots,Q_k$ are constructed,
and array $B$ is partitioned into blocks of size $s$.
For each $i \in \{0, \ldots, n\}$ such that $i \bmod s = 0$, 
construct an array $F_i[1:m]$ such that for each $x$, $F_i[x]$ stores
the largest $j \leq n$ such that the mode of $B[i:j]$ has frequency at most $x$;
a corresponding mode is also stored.
A query range $B[i:j]$ is divided into prefix, span, and suffix subarrays
as before. 
Let $p = s \lceil i / s \rceil$ denote the index of the first element 
of the span.
Using the technique of Greve et al.\ \cite{greve2010},
a mode of the span and its frequency
are computed by finding the successor of $j$ in $F_i$;
this can be achieved in $O(\log\log n)$ time by an $O(n)$-space data structure 
(e.g., a van Emde Boas tree 
\cite{vanEmdeBoas1975,vanEmdeBoas1976,vanEmdeBoas1977}
or a y-fast trie \cite{willard1983}).
By Lemma~\ref{lem:minFreqQuery},
determining whether the frequency 
of an element in the prefix or suffix exceeds that of the mode of the span
requires only constant time per element, or $O(s)$ total time.
The resulting worst-case query time is $O(s + \log\log n)$ 
using $O(n + n \cdot m / s)$ space.
Choosing $s\in\Theta(m)$ gives $O(n)$ space and $O(m + \log\log n)$ query time.

%%%%%%%%%%%%%%%%%%%%%%%%%%%%%%%%%%%%%%%%%%%%
% SUBSECTION
\subsection{Counting Method: $O(|j-i|)$ Query Time and $O(n)$ Space}
\label{sec:dataStructures.j-i}
\noindent
We briefly describe an $O(|j-i|)$-time and $O(n)$-space data structure.
No actual precomputation is necessary
other than constructing the array $B$, finding $k$, 
and initializing a frequency table $C[1:k]$ to zero,
all of which can be achieved in $O(n \log k)$ precomputation time.
%Thus preprocessing time required is $O(k)$ 
%(or $O(n\log k)$ time if $k$ or $B$ must be computed).
This algorithm is similar to counting sort: 
compute a frequency table for $B[i:j]$ stored in $C[1:k]$,
%count the frequency of each element in $A[i:j]$ using array $C[1:k]$,
then identify a maximum element in $C[1:k]$.
When computing the maximum, the running time is bounded to $O(|j-i|)$
by only examining indices in $C$ that correspond to elements in $B[i:j]$
(these are exactly the elements of $C$ that have non-zero values).
%that have been modified during the counting phase. 
This procedure is repeated after identifying the maximum 
to reset $C[1:k]$ to zero.
%
%A query on the range $B[i:j]$ is performed as follows:
%\begin{enumerate}
%\item For each $x \in \{ i, \ldots, j \}$, increment $C[B[x]]$.
%Upon completing this step,
%$C$ is a frequency table for $B[i:j]$.
%\item Find a maximum value in $C$.
%Only non-zero values need to be considered; i.e., 
%check whether $C[B[x]]$ is a maximum for each $x \in \{ i, \ldots, j \}$. 
%If $x'$ is an index that maximizes $C[B[x']]$, then $B[x']$ is a mode 
%of $B[i:j]$.
%\item For each $x \in \{ i, \ldots, j \}$, set $C[B[x]]$ to zero.
%This returns the data structure to its initial state before the next query.
%\end{enumerate}
%
%%%%%%%%%%%%%%%%%%%%%%%%%%%%%%%%%%%%%%%%%%%%
%\subsubsubsection{Storage Space and Query Time}
Each step requires $\Theta(|j-i|)$ time
%Only the arrays $A$, $B$, and $C$ are stored.
and the total space required by the data structure is $O(n)$.

%%% remove following text
%\iffalse

%%%%%%%%%%%%%%%%%%%%%%%%%%%%%%%%%%%%%%%%%%%%
% SECTION
\section{Higher Dimensions}
\label{sec:higherDim}

\noindent
A natural question is whether our results for one-dimensional range mode query 
extend to arbitrary dimensions.  
The array $B[1:n]$ is replaced by a $d$-dimensional array 
$B[1:n_1, \ldots, 1:n_d]$,
containing $n$ elements in total with dimensionality $n_1, \ldots, n_d$,
where $n = n_1 \times \cdots \times n_d$.
Within Section~\ref{sec:higherDim} we refer to a $d$-dimensional tuple 
(e.g., $\vec{i}=[i_1,\ldots,i_d])$ as an array index (e.g., $B[\vec{i}]$).  
%To facilitate discussion about ranges, 
We say a tuple $\vec{i}$ dominates another tuple $\vec{j}$ 
if and only if $i_t \leq j_t$ for all $t \in\{1,\ldots, d\}$.  
%In tuple notation, 
We denote the input array as $B[\vec{1}:\vec{n}]$,
where $\vec{n} = [n_1, \ldots, n_d]$.
%Ranges are no longer over an interval $[i:j]$ of indices, rather, 
A range is defined over a $d$-dimensional rectangle of indices, uniquely
determined by two indices, $[\vec{i}:\vec{j}]$, where $\vec{i}\leq \vec{j}$.

A key element of our one-dimensional data structures is the use of frequency 
tables. 
%To extend this to higher dimensions we need to compute the frequency of an 
%element found at index $\vec{x}$ over the interval $[\vec{a}:\vec{b}]$.  
%Assume the array element $B[\vec{x}]$ contains the rank of array 
%element $A[\vec{x}]$.  
In $d$ dimensions, array $C[1:k]$ is a frequency table
for $B[\vec{a}:\vec{b}]$ if, for each $i\in \{1,\ldots, k\}$,
$C[i]$ stores the number of occurrences of element 
$B[\vec{x}] = i$ in $B[\vec{a}:\vec{b}]$.
%Thus, $C[B[\vec{x}]]$ corresponds to the frequency of $A[\vec{x}]$ 
%in $A[\vec{a}:\vec{b}]$.
Unlike the one-dimensional case, 
if $C_{\vec{i}}[1:k]$ is a frequency table for $B[\vec{1}:\vec{i}]$ and
$C_{\vec{j}}[1:k]$ is a frequency table for $B[\vec{1}:\vec{j}]$,
then $C_{\vec{j}}[B[\vec{x}]]-C_{\vec{i}}[B[\vec{x}]]$ is not
the frequency of $B[\vec{x}]$ in $B[\vec{i}:\vec{j}]$ in general.
%The difference between single and higher dimensionality is that 
In one dimension, $\vec{i}$ dominates 
all indices that are to be excluded from the count, whereas this is not 
the case in higher dimensions. Instead, the $2^d$ 
corners of the $d$-rectangle $[\vec{i}:\vec{j}]$ can be used 
to compute the frequency table with 
typical inclusion-exclusion rules \cite{dujmovic2009}.  
The result is computed using $2^d$ $d$-directional range counting queries to 
determine the frequency of $B[\vec{x}]$ in $B[\vec{a}:\vec{b}]$.  
In the range searching literature it is typical to assume $d$ to be a small 
known constant and for the corresponding factors of $d$ 
to be omitted from the evaluation of space and time requirements.

%%%%%%%%%%%%%%%%%%%%%%%%%%%%%%%%%%%%%%%%%%%%
% SUBSECTION
\subsubsubsection{Counting Method}
\label{sec:higherDim.Count}
The counting method described in Section~\ref{sec:dataStructures.j-i} 
does not depend on any properties of one-dimensional data and extends 
to $d$-dimensional data and queries. 
%The only change in its description is 
%in the use of tuples instead of integer indices and that the items in
%the query $d$-rectangle $[\vec{i}:\vec{j}]$ must be scanned 
%instead of only a single interval $[i:j]$. 
The query time is directly 
proportional to the cardinality of the query range $[\vec{i}:\vec{j}]$:
$O(\prod_{l=1}^d(j_l-i_l+1) )$.  
Precomputation time, query time, 
and space requirements are analogous to those of 
the one-dimensional data structure.

%%%%%%%%%%%%%%%%%%%%%%%%%%%%%%%%%%%%%%%%%%%%
% SUBSECTION
\subsubsubsection{Sparse Frequency Table Method}
\label{sec:higherDim.Sparse}
We now consider a generalization to $d$ dimensions 
of the sparse frequency table method described in 
Section~\ref{sec:dataStructures.k}.
%An obvious improvement to the counting method described in 
%Section~\ref{sec:higherDim.Count} is to 
As in the one-dimensional data structure, for every $\vec{t} \in T$
we precompute a frequency table
$C_{\vec{t}}[1:k]$ for the range $B[\vec{1}:\vec{t}]$,
where $T \subseteq [\vec{1}:\vec{n}]$ is a fixed subset of indices.
If $T$ is a sparse set whose elements are distributed regularly 
across $[\vec{1},\vec{n}]$, then a frequency table 
for the span can be computed in $O(2^d k)$ time and $O(n)$ space
using the inclusion-exclusion principle.
The remainder of the query algorithm consists 
of examining each index $\vec{w}$ in the enclosing set
$W =  [\vec{i}:\vec{j}] \setminus [\vec{b_i}:\vec{b_j}]$
(known as the suffix and prefix in the one-dimensional case)
and incrementing the corresponding frequency count $C[B[\vec{w}]]$.
Finally, the maximum value of the frequency table $C$ 
determines the frequency of the mode; this maximum is identified in $O(k)$ time.
Therefore, the total query time is $O(2^d k + |W|)$.

The regular positioning of the indices in $T$ forms a $d$-dimensional grid that
divides $B[\vec{1}:\vec{n}]$ evenly into $|T|$ cells, 
each of which is a $d$-rectangle of cardinality $s=n/|T|$. 
Each frequency table has size $k$.
In order for the space occupied by the frequency tables to remain linear
there can be at most $O(n/k)$ such tables (e.g., let $|T| = \lceil n/k \rceil$
and $s = k$).  
We set the width of each cell in the $l$ dimension
to be $O(n_l(k/n)^{\frac{1}{d}})$.  
Observe that $\prod_{l=1}^d n_l(k/n)^{\frac{1}{d}} = k$.  
Since there are $s=k$ items in a cell, the number of items on the 
cell's surface perpendicular to dimension $l$ is 
\[ O\left(\frac{k}{n_l}\left(\frac{n}{k}\right)^{\frac{1}{d}}\right)
=O\left(k^{\frac{d-1}{d}}\frac{n^{\frac{1}{d}}}{n_l}\right) . \]

%We now examine the value $|W|$.
Observe that $|W|$ is at most $s$ times the number of 
cells on the external surfaces of the $d$-rectangle specified by the query 
range $[\vec{i},\vec{j}]$.  The total number of items on the external 
surface perpendicular to some dimension $l\in\{1,\ldots,d\}$ is $O(n/n_l)$.
%within a constant dependent on $d$,
Thus the number of cells on that external surface is 
\[ O\left(\frac{n}{n_l}  
\frac{1}{k^{\frac{d-1}{d}}}\frac{n_l}{n^{\frac{1}{d}}}\right) 
= O\left( \left(\frac{n}{k}\right)^\frac{d-1}{d} \right). \]
Therefore, $|W| \in O(d \cdot k(n/k)^\frac{d-1}{d}) 
=O(d \cdot n^\frac{d-1}{d}k^{\frac{1}{d}})$,
resulting in a total query time of
$O(2^d k + d \cdot n^\frac{d-1}{d}k^{\frac{1}{d}})$.
If $k$ is constant, then the query time can be improved to
$O(2^d k)$ using $O(n \cdot k)$ space 
by including a frequency table for every item in $B$.

%%%%%%%%%%%%%%%%%%%%%%%%%%%%%%%%%%%%%%%%%%%%
% SUBSECTION
\subsubsubsection{Sparse Mode Table Method}
\label{sec:higherDim.SparseMode}
The sparse mode table method described in Section~\ref{sec:dataStructure1}
and the sparse frequency table method 
both specify a subset $T$ of indices positioned at regular intervals
for which any pair determines a span within the array $B$.  
Instead of storing frequencies for all elements in $D$, however, 
the sparse mode table method stores 
a precomputed mode of the span between any two indices in $T$.   
The mode of the query range is then found by searching for elements
in the prefix and suffix whose frequency exceeds that of the mode of the span.

This data structure exemplifies the space-time trade-off.
The $O(\sqrt{n})$ query time and $O(n)$ space bounds of
the one-dimensional data structure 
are possible because the cardinality of the prefix and suffix can be kept small
while minimizing the time required to measure 
the frequency of elements in the prefix and suffix.
In particular, the one-dimensional data structure 
supports a constant-time query to determine whether the frequency of
a given element exceeds that of the mode of the span.
This is achieved by referring to the arrays $Q_1, \ldots, Q_k$.
%Determining if a particular element from the enclosing set 
%(i.e., suffix/prefix)
%is such a candidate is done in $O(1)$ time per occurrence in the suffix/prefix 
%using an array of arrays $Q$ (see Figure~\ref{fig:dataStructure1}).
%Unfortunately the time versus space tradeoff is not as effective when this 
%data structure is extended to higher dimensions.
These arrays, however, do not generalize easily to higher dimensions.
%In higher dimensions an equivalent data structure for $Q$ is not strictly 
%available.
A corresponding decision query would be: ``Does element $B[\vec{x}]$ occur at
least $m$ times in the block $B[\vec{i}:\vec{j}]$?''
Replacing the arrays $Q_1, \ldots, Q_k$ with orthogonal range counting 
data structures answers the query: ``How frequently does element $B[\vec{x}]$ 
occur in the block $B[\vec{i}:\vec{j}]$?''
A range counting query computed using $kd$-trees 
gives a linear-space data structure with 
$O(|Q[B[\vec{x}]]|^{1-\frac{1}{d}})$ query time \cite{lee1977}.  
Bentley and Mauer \cite{bentley1980} describe a linear-space 
data structure with a faster query time of 
$O(|Q[B[\vec{x}]]|^{\epsilon})$ for any fixed $\epsilon<1$,
where the time and space bounds omit constant factors of $\epsilon$.

As in Section~\ref{sec:higherDim.Sparse},
let $W$ denote the enclosing set of indices, 
(i.e., the indices of the query range not contained in the span).  
Let $D_W$ denote the set of distinct elements contained in $W$.
Thus the range mode query time is\footnote{Our data structure
includes $kd$-trees. In the corresponding analysis of Lee and Wong 
\cite{lee1977}, $d$ is assumed to be constant; 
consequently, constants dependent upon $d$ 
do not appear in \eqref{eqn:sparseMode}.}
\begin{equation}
\label{eqn:sparseMode}
O \left( \max\left\{ \sum_{u\in D_W}|Q[u]|^{\frac{d-1}{d}} , |W|\right\} 
\right) 
\subseteq O \left( \max \left\{ n^{\frac{d-1}{d}} , |W| \right\} \right). 
\end{equation}

The arrays $S$ and $S'$ respectively store a mode and frequency
of the span $B[\vec{b_i}:\vec{b_j}]$ 
for all $\{\vec{b_i}, \vec{b_j}\} \subseteq T$.
Maintaining linear space requires 
that $\Theta(|T|) = \Theta(s) = \Theta(\sqrt{n})$.
We set the number of elements per cell in the $l$ dimension to 
be $O(\sqrt{n_l})$. Thus the number of elements on the 
surface of the cell perpendicular to the $l$ dimension is $O(\sqrt{n/n_l})$.
The total number of elements on the external surface perpendicular to some 
dimension $l\in\{1,\ldots,d\}$ is $O(n/n_l)$.
%proportional, within a constant dependent on $d$
Thus the number of cells on the external surface is 
$O((n/n_l)\sqrt{n_l/n}) = O(\sqrt{n/n_l})$.
Therefore, 
\begin{equation}
\label{eqn:cardinalityW}
|W| \in O\left( n \sum_{l=1}^d \frac{1}{\sqrt{n_l}} \right) .
\end{equation}
If all values $n_l$ are equal, then
\eqref{eqn:cardinalityW} simplifies to $O(d \cdot n^{1-\frac{1}{2d}})$.

%%% end of removed text
%\fi

%%%%%%%%%%%%%%%%%%%%%%%%%%%%%%%%%%%%%%%%%%%%
% SECTION
\section{Discussion and Directions for Future Research}
\label{sec:discussion}

%%%%%%%%%%%%%%%%%%%%%%%%%%%%%%%%%%%%%%%%%%%%
% SUBSECTION
%\subsubsubsectionB{Higher Dimensions}
%\label{sec:higherDim}

%%%%%%%%%%%%%%%%%%%%%%%%%%%%%%%%%%%%%%%%%%%%
% SUBSECTION
\subsubsubsectionB{Generalizing Mode}
\label{sec:discussion.generalization}
The sparse frequency table and counting methods described 
in Sections~\ref{sec:dataStructures.k} and~\ref{sec:dataStructures.j-i},
respectively, can be generalized to return
the $x$th most frequently occurring element in the query range $A[i:j]$ for any 
$x \in \{1, \ldots, k\}$ by employing a linear-time 
($O(\min\{k, |j-i|\})$ time) selection algorithm
to find the $x$th largest element in the frequency table for $A[i:j]$.
%This requires collecting the non-zero elements of array $C$
%in the first algorithm.
Due to its dependence on precomputed modes stored in array $S$,
an analogous generalization seems unlikely 
without a significant increase in space for the sparse mode table method
described in Section~\ref{sec:dataStructure1}.
%Greve et al.\ \cite{greve2010} examined related problems recently.
%An interesting open problem is to construct a linear-space data structure
%for identifying 
%the $x$th most frequently occuring element in the query range $A[i:j]$ 
%with $O(\sqrt{n})$ worst-case query time.

\begin{openProb}
\label{open:generalization}
Given a list of $A[1:n]$ of $n$ items, 
construct an $O(n)$-space data structure for identifying 
the $x$th most frequently occurring element in the range $A[i:j]$ 
with $O(\sqrt{n})$ query time, 
where $i$, $j$, and $x$ are provided at query time.
\end{openProb}

%%%%%%%%%%%%%%%%%%%%%%%%%%%%%%%%%%%%%%%%%%%%
% SUBSECTION
\subsubsubsectionB{Dynamic Range Mode Query}
\label{sec:discussion.dynamic}
Prior discussion has been restricted to static data structures 
for range mode query.
Dynamically updating the list of items 
is a natural operation: $A[i] \leftarrow x$.
Unlike the range median query problem for which dynamic data structures
exist \cite{brodal2009,brodal2010,gfeller2009,jorgensen2010},
none of the previous data structures for range mode query 
\cite{bose2005,krizanc2005,greve2010,petersen2008,petersen2009} support 
efficient updates. We briefly discuss 
some of the challenges of making our data structures dynamic.

Both the sparse frequency table and counting methods
described in Sections~\ref{sec:dataStructures.k}
and~\ref{sec:dataStructures.j-i}, respectively,
permit straightforward constant-time updates
when the set of distinct elements, $D$, remains unchanged.
Updates that modify $D$, however, require careful consideration.
A key issue in defining dynamic data structures analogous to the static
data structures described in this paper
%Section~\ref{sec:dataStructures}
is to generalize the mapping defined by array $B$ 
(see Section~\ref{sec:dataStructure1})  
to support efficient updates.
We have preliminary results demonstrating that such updates
are possible for implementing a dynamic version of the counting method.
As the data structure for the sparse frequency method is currently specified, 
however, updates that modify $D$ require
$\Theta(n)$ time in the worst case.
The sparse mode table method described in Section~\ref{sec:dataStructure1}
does not suggest itself as a good candidate for efficient updates.
In particular, the table $S$ requires $\Theta(n)$ updates in the worst case,
even if $D$ remains unchanged.
%For example, consider an alternating sequence over
%the binary universe $\{0,1\}$.
%Suppose the two middle items are changed to 1; 
%the unique mode of all sequences of
%blocks that span the two modified items is now 1.
%Suppose the two middle items are then changed to 0; the mode of all sequences 
%of blocks that span the two modified items changes to 0.
%Thus, updating $O(1)$ items in array $A$ requires $\Omega(t^2)$
%updates to entries in table $S$ in the worst case
%(to achieve $O(\sqrt{n})$ query time, our static data structure
%requires $t^2 \in \Omega(n)$).
Also challenging is the problem of updating the arrays $Q_1, \ldots, Q_k$.
Each set $Q_x$ is stored as a sorted array to enable direct indexing,
resulting in $\Theta(n)$ update time in the worst case.
%Updating a single item in array $A$ can result
%in $\Theta(n)$ changes to two arrays $Q_x$ and $Q_{x'}$.
Thus, the problem of
defining an efficient dynamic range mode query data structure remains open.

\begin{openProb}
\label{open:dynamic}
Given an array $A[1:n]$ of $n$ items, construct a dynamic data structure
that supports efficient range mode queries and updates.
\end{openProb}

%%%%%%%%%%%%%%%%%%%%%%%%%%%%%%%%%%%%%%%%%%%%
% SUBSECTION
\subsubsubsectionB{Geometric Range Mode Query}
\label{sec:discussion.rangeSearch}
The range mode problem has a natural definition in Euclidean space:

\begin{openProb}
\label{open:continuousRangeMode}
Given a multiset $P$ of $n$ points in $\reals^d$,
construct a data structure 
to support queries that return a mode of $P \cap R$ 
for an arbitrary (orthogonal) query range $R \subseteq \reals^d$.
What is the time complexity of such a range query for a given space bound?
\end{openProb}
%\begin{openProb}
%\label{open:continuousRangeMode2}
%Given a set $P$ of $n$ points in $\reals^d$, each of which is assigned
%a colour from the set $\{ 1, \ldots , k \}$,
%construct a data structure 
%to support queries that return a mode of the multiset determined 
%by the colours of points in $P \cap R$,
%where $R \subseteq \reals^d$ is an arbitrary (orthogonal) query range.
%What is the time complexity such a range query for a given space bound?
%\end{openProb}

Interpreted differently, an instance of Problem \ref{open:continuousRangeMode} 
is a set of points $P' \subseteq \reals^d$, 
such that each point $p \in P'$ is assigned a colour.
In this case, the mode of $R \cap P'$ is the most frequently occuring colour
in the query region. 
As discussed in Section~\ref{sec:relatedWork.continuous},
when $d=1$, this problem reduces to range mode query on an array.
When $d \geq 2$, however,
solution techniques tend to differ extensively for
range searching problems set in continuous Euclidean space
versus those restricted to array input.

A range reporting query can be combined with a mode-finding algorithm 
(e.g., the counting method described in Section~\ref{sec:dataStructures.j-i})
to identify the multiset of points within the query range 
and then compute its mode.
% (this second step is independent of geometry).
%This solution, however, does not precompute any information specific
%to the mode, other than counting the number of distinct points in $P$.
Such a solution requires enumerating all elements in the query range,
possibly resulting in poor query time (e.g., when $|R \cap P| \in \Theta(|P|)$).
A more ingenious solution might reduce query time 
by avoiding the use of a range report query.
Other than a basic combination approach such as that described above,
%of combining a range reporting query 
%%to identify the multiset of points within the query range 
%with a mode algorithm without additional preprocessing
the range mode query problem in the continuous setting remains open.

%%%%%%%%%%%%%%%%%%%%%%%%%%%%%%%%%%%%%%%%%%%%
% SUBSECTION
\subsubsubsection{Lower Bounds}
\label{sec:discussion.lowerBounds}
Recently, Greve et al.\ \cite{greve2010} showed that any data structure
that uses $s$ memory cells of $w$ bits requires
$\Omega(\log n / \log(s\cdot w/n) )$ time to answer a range mode query.
For linear-space data structures in the RAM model, 
$s \cdot w \in \Theta(n \log n)$, 
corresponding to a lower bound of $\Omega(\log n / \log\log n)$ query time.
Other than the bound of Greve et al.\ 
and the lower bounds on the problem of computing a mode of 
a multiset (see Section~\ref{sec:relatedWork.mode}), 
little is known regarding non-trivial lower bounds 
for the time complexity of the range mode query problem.
In particular, it is unknown whether there exists a linear-space data structure
that supports $o(\sqrt{n})$ query time.

\begin{openProb}
\label{open:lowerBound}
Identify a function $f(n)$
such that any $O(n)$-space data structure that supports range mode query
on an array of $n$ items requires $\Omega(f(n))$ query time in the
worst case, where $f(n) \in \omega(\log n/\log\log n)$,
or provide an $O(n)$-space data structure that 
supports $O(\log n/\log\log n)$-time queries.
\end{openProb}

The corresponding question for range selection query was recently
solved by J{\o}rgensen and Larsen \cite{jorgensen2011} who showed
a lower bound of $\Omega(\log r / \log\log n)$ and 
a linear-space data structure with $O(\log r / \log\log n + \log\log n)$ 
query time, where $r$ denotes the rank of the selection query.

%%%%%%%%%%%%%%%%%%%%%%%%%%%%%%%%%%%%%%%%%%%%
% ACKNOWLEDGEMENTS
%\section*{Acknowledgements}

\subsubsubsection{Acknowledgements}
The authors thank
Peyman Afshani, Timothy Chan,
Francisco Claude, Meng He, Ian Munro, Patrick Nicholson, 
Matthew Skala, and Norbert Zeh for discussing various topics related
to range searching.
%, including
%wavelet trees, batched predecessor queries, 
%range counting in higher dimensions, and majority range queries.

%The authors thank Francisco Claude with whom
%they discussed wavelet trees and batched predecessor queries,
%Peyman Afshani and Norbert Zeh 
%with whom they discussed range counting in higher dimensions,
%and Ian Munro, Patrick Nicholson, Matt Skala, and Meng He
%with whom they discussed majority range queries.

%%%%%%%%%%%%%%%%%%%%%%%%%%%%%%%%%%%%%%%%%%%%
% BIBLIOGRAPHY

%\footnotesize

\bibliographystyle{plain}
\bibliography{rangeModeQueryNotes}

\begin{thebibliography}{10}

\bibitem{agarwal2004}
P.~K. Agarwal.
\newblock Range searching.
\newblock In J.~Goodman and J.~O'Rourke, editors, {\em Handbook of Discrete and
  Computational Geometry}, pages 809--837. CRC Press, New York, 2nd edition,
  2004.

\bibitem{alon1987}
N.~Alon and B.~Schieber.
\newblock Optimal preprocessing for answering on-line product queries.
\newblock Technical Report 71/87, Tel-Aviv University, 1987.

\bibitem{amir2007}
A.~Amir, J.~Fischer, and M.~Lewenstein.
\newblock Two-dimensional range minimum queries.
\newblock In {\em Proceedings of the Symposium on Combinatorial Pattern
  Matching (CPM)}, volume 4580 of {\em Lecture Notes in Computer Science},
  pages 286--294. Springer, 2007.

\bibitem{bender2000}
M.~A. Bender and M.~Farach-Colton.
\newblock The {LCA} problem revisited.
\newblock In {\em Proceedings of the Latin American Theoretical Informatics
  Symposium (LATIN)}, volume 1776 of {\em Lecture Notes in Computer Science},
  pages 88--94. Springer, 2000.

\bibitem{bentley1980}
J.~L. Bentley and H.~A. Maurer.
\newblock Efficient worst-case data structures for range searching.
\newblock {\em Acta Informatica}, 13(2):155--168, 1980.

\bibitem{berkman1989}
O.~Berkman, D.~Breslauer, Z.~Galil, B.~Schieber, and U.~Vishkin.
\newblock Highly parallelizable problems.
\newblock In {\em Proceedings of the {ACM} Symposium on the Theory of Computing
  (STOC)}, pages 309--319, 1989.

\bibitem{bose2005}
P.~Bose, E.~Kranakis, P.~Morin, and Y.~Tang.
\newblock Approximate range mode and range median queries.
\newblock In {\em Proceedings of the International Symposium on Theoretical
  Aspects of Computer Science (STACS)}, volume 3404 of {\em Lecture Notes in
  Computer Science}, pages 377--388. Springer, 2005.

\bibitem{brodal2010b}
G.~S. Brodal, P.~Davoodi, and S.~S. Rao.
\newblock On space efficient two dimensional range minimum data structures.
\newblock In {\em Proceedings of the European Symposium on Algorithms (ESA)},
  volume 6346/6347 of {\em Lecture Notes in Computer Science}. Springer, 2010.

\bibitem{brodal2010}
G.~S. Brodal, B.~Gfeller, A.~G. J{\o}rgensen, and P.~Sanders.
\newblock Towards optimal range medians.
\newblock {\em Theoretical Computer Science}, 2011.
\newblock In press.

\bibitem{brodal2009}
G.~S. Brodal and A.~G. J{\o}rgensen.
\newblock Data structures for range median queries.
\newblock In {\em Proceedings of the International Symposium on Algorithms and
  Computation (ISAAC)}, volume 5878 of {\em Lecture Notes in Computer Science},
  pages 822--831. Springer, 2009.

\bibitem{chan2010}
T.~Chan and M.~P\u{a}tra\c{s}cu.
\newblock Counting inversions, offline orthogonal range counting, and related
  problems.
\newblock In {\em Proceedings of the {ACM-SIAM} Symposium on Discrete
  Algorithms (SODA)}, pages 161--173, 2010.

\bibitem{chazelle1989}
B.~Chazelle and B.~Rosenberg.
\newblock Computing partial sums in multidimensional arrays.
\newblock In {\em Proceedings of the {ACM} Symposium on Computational Geometry
  (SoCG)}, pages 131--139, 1989.

\bibitem{demaine2009}
E.~D. Demaine, G.~M. Landau, and O.~Weimann.
\newblock On {Cartesian} trees and range minimum queries.
\newblock In {\em Proceedings of the International Colloquium on Automata,
  Languages, and Programming (ICALP)}, volume 5555 of {\em Lecture Notes in
  Computer Science}, pages 341--353. Springer, 2009.

\bibitem{dujmovic2009}
V.~Dujmovi\'c, J.~Howat, and P.~Morin.
\newblock Biased range trees.
\newblock In {\em Proceedings of the {ACM-SIAM} Symposium on Discrete
  Algorithms (SODA)}, pages 486--495, 2009.

\bibitem{vanEmdeBoas1975}
P.~van Emde~Boas.
\newblock Preserving order in a forest in less than logarithmic time.
\newblock In {\em Proceedings of the {IEEE} Symposium on Foundations of
  Computer Science}, pages 75--84, 1975.

\bibitem{vanEmdeBoas1977}
P.~van Emde~Boas.
\newblock Preserving order in a forest in less than logarithmic time and linear
  space.
\newblock {\em Information Processing Letters}, 6(3):80--82, 1977.

\bibitem{vanEmdeBoas1976}
P.~van Emde~Boas, R.~Kaas, and E.~Zijlstra.
\newblock Design and implementation of an efficient priority queue.
\newblock {\em Mathematical Systems Theory}, 10:99--127, 1976.

\bibitem{fischer2010}
J.~Fischer.
\newblock Optimal succinctness for range minimum queries.
\newblock In {\em Proceedings of the Latin American Theoretical Informatics
  Symposium (LATIN)}, volume 6034 of {\em Lecture Notes in Computer Science},
  pages 158--169. Springer, 2010.

\bibitem{fischer2006}
J.~Fischer and V.~Heun.
\newblock Theoretical and practical improvements on the {RMQ}-problem, with
  applications to {LCA} and {LCE}.
\newblock In {\em Proceedings of the Symposium on Combinatorial Pattern
  Matching (CPM)}, volume 4009 of {\em Lecture Notes in Computer Science},
  pages 36--48. Springer, 2006.

\bibitem{fischer2007}
J.~Fischer and V.~Heun.
\newblock A new succinct representation of {RMQ}-information and improvements
  in the enhanced suffix array.
\newblock In {\em Proceedings of the International Symposium on Combinatorics,
  Algorithms, Probabilistic and Experimental Methodologies (ESCAPE)}, volume
  4614 of {\em Lecture Notes in Computer Science}, pages 459--470. Springer,
  2007.

\bibitem{fischer2010b}
J.~Fischer and V.~Heun.
\newblock Finding range minima in the middle: Approximations and applications.
\newblock {\em Mathematics in Computer Science}, 3(1):17--30, 2010.

\bibitem{gabow1984}
H.~N. Gabow, J.~L. Bentley, and R.~E. Tarjan.
\newblock Scaling and related techniques for geometry problems.
\newblock In {\em Proceedings of the {ACM} Symposium on the Theory of Computing
  (STOC)}, pages 135--143, 1984.

\bibitem{gagie2009}
T.~Gagie, S.~J. Puglisi, and A.~Turpin.
\newblock Range quantile queries: Another virtue of wavelet trees.
\newblock In {\em Proceedings of the String Processing and Information
  Retrieval Symposium (SPIRE)}, volume 5721 of {\em Lecture Notes in Computer
  Science}, pages 1--6. Springer, 2009.

\bibitem{gfeller2009}
B.~Gfeller and P.~Sanders.
\newblock Towards optimal range medians.
\newblock In {\em Proceedings of the International Colloquium on Automata,
  Languages, and Programming (ICALP)}, volume 5555 of {\em Lecture Notes in
  Computer Science}, pages 475--486. Springer, 2009.

\bibitem{greve2010}
M.~Greve, A.~G. J{\o}rgensen, K.~D. Larsen, and J.~Truelsen.
\newblock Cell probe lower bounds and approximations for range mode.
\newblock In {\em Proceedings of the International Colloquium on Automata,
  Languages, and Programming (ICALP)}, volume 6198 of {\em Lecture Notes in
  Computer Science}, pages 605--616. Springer, 2010.

\bibitem{har-peled2008}
S.~Har-Peled and S.~Muthukrishnan.
\newblock Range medians.
\newblock In {\em Proceedings of the European Symposium on Algorithms (ESA)},
  volume 5193 of {\em Lecture Notes in Computer Science}, pages 503--514.
  Springer, 2008.

\bibitem{jaja2004}
J.~{J\'aJ\'a}, C.~W. Mortensen, and Q.~Shi.
\newblock Space-efficient and fast algorithms for multidimensional dominance
  reporting and counting.
\newblock In {\em Proceedings of the International Symposium on Algorithms and
  Computation (ISAAC)}, volume 3341 of {\em Lecture Notes in Computer Science},
  pages 558--568. Springer, 2004.

\bibitem{jorgensen2010}
A.~G. J{\o}rgensen.
\newblock {\em Data Structures: Sequence Problems, Range Queries, and Fault
  Tolerance}.
\newblock PhD thesis, Aarhus University, 2010.

\bibitem{jorgensen2011}
A.~G. J{\o}rgensen and K.~D. Larsen.
\newblock Range selection and median: Tight cell probe lower bounds and
  adaptive data structures.
\newblock In {\em Proceedings of the {ACM-SIAM} Symposium on Discrete
  Algorithms (SODA)}, 2011.
\newblock To appear.

\bibitem{krizanc2005}
D.~Krizanc, P.~Morin, and M.~Smid.
\newblock Range mode and range median queries on lists and trees.
\newblock {\em Nordic Journal of Computing}, 12:1--17, 2005.

\bibitem{lee1977}
D.~T. Lee and C.~K. Wong.
\newblock Worst-case analysis for region and partial region searches in
  multidimensional binary search trees and balanced quad trees.
\newblock {\em Acta Informatica}, 9(1):23--29, 1977.

\bibitem{munro1976}
J.~I. Munro and M.~Spira.
\newblock Sorting and searching in multisets.
\newblock {\em {SIAM} Journal on Computing}, 5(1):1--8, 1976.

\bibitem{petersen2008}
H.~Petersen.
\newblock Improved bounds for range mode and range median queries.
\newblock In {\em Proceedings of the Conference on Current Trends in Theory and
  Practice of Computer Science (SOFSEM)}, volume 4910 of {\em Lecture Notes in
  Computer Science}, pages 418--423. Springer, 2008.

\bibitem{petersen2009}
H.~Petersen and S.~Grabowski.
\newblock Range mode and range median queries in constant time and
  sub-quadratic space.
\newblock {\em Information Processing Letters}, 109:225--228, 2009.

\bibitem{poon2003}
C.~K. Poon.
\newblock Optimal range max datacub for fixed dimensions.
\newblock In {\em Proceedings of the International Conference on Database
  Theory (ICDT)}, volume 2572 of {\em Lecture Notes in Computer Science}, pages
  158--172. Springer, 2003.

\bibitem{skiena2008}
S.~Skiena.
\newblock {\em The Algorithm Design Manual}.
\newblock Springer, 2nd edition, 2008.

\bibitem{willard1983}
D.~E. Willard.
\newblock Log-logarithmic worst-case range queries are possible in space
  {$\Theta(N)$}.
\newblock {\em Information Processing Letters}, 17:81--84, 1983.

\bibitem{yao1982}
A.~C. Yao.
\newblock Space-time tradeoff for answering range queries.
\newblock In {\em Proceedings of the {ACM} Symposium on the Theory of Computing
  (STOC)}, pages 128--136, 1982.

\bibitem{yao1985}
A.~C. Yao.
\newblock On the complexity of maintaining partial sums.
\newblock {\em SIAM Journal on Computing}, 14:277--288, 1985.

\bibitem{yuan2010}
H.~Yuan and M.~J. Atallah.
\newblock Data structures for range minimum queries.
\newblock In {\em Proceedings of the {ACM-SIAM} Symposium on Discrete
  Algorithms (SODA)}, pages 150--160, 2010.

\end{thebibliography}

\end{document}